\documentclass[amsmath,amssymb,floatfix,aps,prl,reprint,twocolumn,superscriptaddress]{revtex4-1}

\usepackage{graphicx}
\usepackage{dcolumn}
\usepackage{bm}
\usepackage{amsmath}
\usepackage{hyperref}

\graphicspath{ {./} }

\begin{document}

\title{Charge Transport in Organic Molecular Semiconductors from First Principles:\\ The Band-Like Hole Mobility in Naphthalene Crystal}

\author{Nien-En Lee}
\affiliation{Department of Applied Physics and Materials Science, Steele Laboratory, California Institute of Technology, Pasadena, California 91125, USA}
\affiliation{Department of Physics, California Institute of Technology, Pasadena, California 91125, USA}

\author{Jin-Jian Zhou}
\affiliation{Department of Applied Physics and Materials Science, Steele Laboratory, California Institute of Technology, Pasadena, California 91125, USA}

\author{Luis A. Agapito}
\affiliation{Department of Applied Physics and Materials Science, Steele Laboratory, California Institute of Technology, Pasadena, California 91125, USA}

\author{Marco Bernardi}
\affiliation{Department of Applied Physics and Materials Science, Steele Laboratory, California Institute of Technology, Pasadena, California 91125, USA}

\date{\today}

\begin{abstract}
\noindent
Predicting charge transport in organic molecular crystals is notoriously challenging. Carrier \mbox{mobility} calculations in organic semiconductors are dominated by quantum chemistry methods based on charge hopping, which are laborious and only moderately accurate. 
We compute from first principles the electron-phonon scattering and the phonon-limited hole mobility of naphthalene crystal in the framework of \textit{ab initio} band theory. 
Our calculations combine GW electronic bandstructures, \textit{ab initio} electron-phonon scattering, and the Boltzmann transport equation. The calculated hole mobility is in very good agreement with experiment between 100$-$300 K, and we can predict its temperature dependence with high accuracy. We show that scattering between inter-molecular phonons and holes regulates the mobility, though intra-molecular phonons possess the strongest coupling with holes. We revisit the common belief that only rigid molecular motions affect carrier dynamics in organic molecular crystals. Our work provides a quantitative and rigorous framework to compute charge transport in organic crystals, and is a first step toward reconciling band theory and carrier hopping computational methods.
\end{abstract} 
\pacs{}

\maketitle
\section{Introduction}
\vspace{-10pt}
Organic molecular crystals are broadly relevant to solid state physics. Their electronic properties range from conducting to insulating, and they can exhibit anisotropic electrical and optical properties, ferroelectricity, magnetism, and superconductivity. 
Organic semiconductors are lead candidates for novel optoelectronics and spintronics applications \cite{Muccini2006, Taliani2009}. Crystals like pentacene and rubrene are already widely used in organic field-effect transistors and light-emitting devices \cite{Bao2006, Someya2010, Arakawa2003}.\\
\indent
Yet, in most organic crystals the nature and transport mechanisms of charge carriers remain unclear. Possible charge transport regimes include polaron charge hopping, band transport, and intermediate regimes, each leading to a peculiar temperature dependence of the mobility. 
Even in the same organic crystal, electrons and holes can behave differently. An example is naphthalene, where hole carriers display band-like transport with a power-law temperature dependence of the mobility \cite{Karl1985}, though electron transport in the out-of-plane direction is polaronic and nearly temperature independent \cite{McGhie1978}.\\ 
\indent
Several approaches have been proposed to compute charge transport in organic crystals \cite{BlumbergerReview}. 
Recent calculations favor either quantum chemistry methods based on hopping of localized charge carriers \cite{BlumbergerReview, Negri2014, Troisi2007, Tully1990, Beljonne2013_02, Orlandi2006, Ciuchi2016, Troisi2017}, or somewhat less extensively polaron theories \cite{Munn1980, Duke1989, Bobbert2004_01, Bobbert2004_02, Hannewald2009, Hannewald2010}. 
%
Charge hopping calculations have provided remarkable insight into charge transport in molecular crystals \cite{BlumbergerReview, Negri2014, Troisi2007, Tully1990, Beljonne2013_02, Orlandi2006, Ciuchi2016, Troisi2017}. However, they are laborious, and are not based on rigorous condensed matter theory. They require large molecular dynamics or Monte Carlo simulations, rely on semiempirical charge transfer models based on Marcus theory, and include the temperature dependence of charge transport only approximately, typically using the Einstein diffusion formula. A common assumption is also that only rigid molecular motions affect the rate of carrier hopping, and therefore charge transport. 
The accuracy of the charge hopping approaches is limited $-$ the best calculations yield mobility values 3$-$4 times greater than experiment \cite{Negri2014,Troisi2007}, though order-of-magnitude discrepancies between computed and measured mobilities are more common \cite{BlumbergerReview}.\\ 
%
%
%
\begin{figure}[!b]
\vspace{-1.0\baselineskip}
\includegraphics[clip=left botm right top, width=0.45\textwidth]{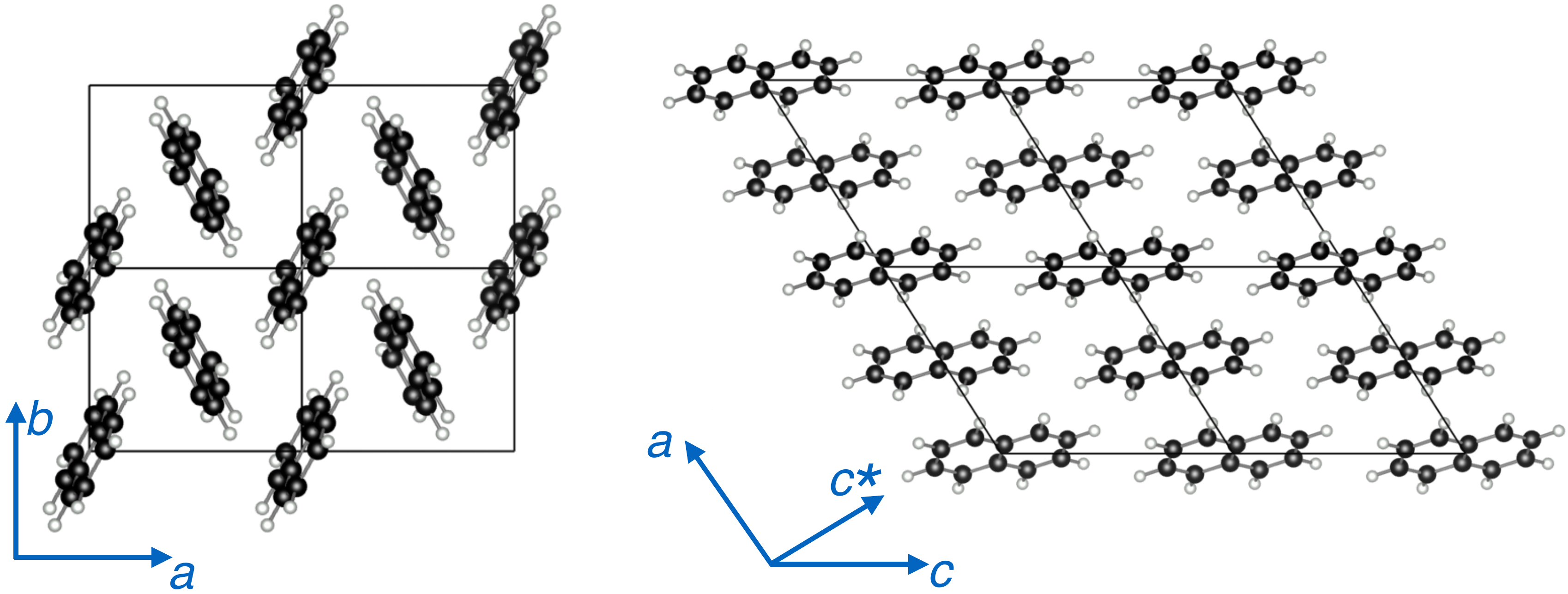}
\caption{The monoclinic crystal structure of naphthalene, with two molecules in the unit cell. The molecules are arranged in a herringbone pattern in the \textit{ab} planes (left), which are stacked in the \textit{c} crystallographic direction (right). The \textit{c}* direction normal to the \textit{ab} plane is also shown.} 
\label{fig:crystal_structure}
\end{figure}
\indent
%
%
To date, only few works have employed band theory to compute charge transport in organic crystals \cite{Shuai2007, Shuai2009, Bredas2003, Kenkre2003}, despite experimental \cite{Hegmann2006, Ramirez2006, Basov2007, Ishii2010} and theoretical \cite{Itabashi2013} evidence of band-like transport in tetracene, rubrene, naphthalene and other organic semiconductors. 
Methods combining band theory and many-body perturbation theory have been recently employed to accurately compute electron-phonon (e-ph) scattering and charge transport, for now in simple inorganic materials with a handful of atoms in the unit cell \cite{Bernardi-noble, Zhou2016, Zhou2017, Bernardi-review}. Due to computational cost, these calculations have not yet been applied to organic crystals with tens of atoms in the unit cell. 
%
%
%
\textit{Ab initio} studies of e-ph coupling in organic crystals exist \cite{FMauri2012,Venuti2010,Stojanovic2012}, 
but charge transport, which requires more elaborate workflows \cite{Zhou2016}, has not yet been investigated within this framework.\\
\indent  
Here we compute from first principles the band-like hole mobility of naphthalene crystal, a material with 36 atoms in the unit cell (see Fig. \ref{fig:crystal_structure}). 
The computed mobility is within a factor of 3$-$4 of experiment, and we can accurately predict its temperature dependence between 100$-$300 K. 
For organic semiconductors, these results are a rare case of very good quantitative agreement with experiment $-$ the accuracy on the mobility is on par with the best charge hopping calculations, and we make an order of magnitude improvement over previous \textit{ab initio} mobility calculations in organic crystals using band theory \cite{Shuai2007,Shuai2009}. 
We show that inter-molecular phonons (i.e., rigid molecular motions) regulate the mobility 
due to a large phase space for scattering holes with energy close to the band edge. Yet, contrary to common notions, intra-molecular phonons exhibit the strongest coupling with holes. 
Our work reconciles the tenet of charge hopping methods that inter-molecular phonons control the mobility with the many-body theory perspective, which treats carrier scattering in terms of phonon absorption and emission events.\\
%
%
\section{methods}
\vspace{-10pt}
\indent 
We carry out density functional theory (DFT) calculations using the {\sc Quantum ESPRESSO} code \cite{QE-2009} with a plane-wave basis set. We employ the Perdew-Burke-Ernzerhof generalized gradient approximation \cite{PBE1996} and norm-conserving pseudopotentials \cite{NormCon1991} from Pseudo Dojo \cite{[http://www.pseudo-dojo.org/\\]Lejaeghereaad3000}. A kinetic energy cutoff of 90 Ry and $4 \times 4 \times 4$ \textbf{k}-point grids are used in all DFT calculations. 
Thermal expansion is taken into account by employing, in separate calculations, lattice constants \cite{Dunitz1982} and atomic positions \cite{LeBail2012,LeBail2009} taken from experiment at four different temperatures of 100, 160, 220, and 300 K. All calculations listed below are repeated separately at these four temperatures. 
The Grimme van der Waals (vdW) correction \cite{Grimme2006,Vittadina2009} is included during structural relaxation. To obtain accurate electronic bandstructures \cite{Neaton2016_02}, we carry out GW calculations using the YAMBO code \cite{Varsano2009}, and obtain the G$_0\!$W$_0$ self-energy using 500 bands in the polarization function and a cutoff of 10 Ry in the dielectric screening. Wannier90 \cite{Marzari2014} is employed to interpolate the bandstructure, using \textit{ab initio} molecular orbitals \cite{Agapito2016} as initial guesses.\\
%
%
\indent 
Phonon dispersions are computed with density functional perturbation theory (DFPT) \cite{Giannozzi2001} on a $2 \times 4 \times 2$ \textbf{q}-point grid \cite{Neaton2016}. The e-ph coupling matrix elements $g_{nm\nu}(\textbf{k},\textbf{q})$ on coarse \textbf{k}- and \textbf{q}-point grids \cite{Bernardi-review} are computed using a routine from the EPW code \cite{Giustino2016} and interpolated using Wannier functions \cite{Giustino2007} generated with the Wannier90 code \cite{Marzari2014}. 
Here and in the following, $n$ and $m$ are band indices, $\nu$ labels phonon modes, and $\textbf{k}$ and $\textbf{q}$ are crystal momenta for electrons and phonons, respectively. Our in-house developed code {\sc Perturbo} \cite{PerturboWebsite} is employed to interpolate the e-ph matrix elements on fine grids with up to $60 \times 60 \times 60$ \textbf{k}-points and $10^5$ random \textbf{q}-points, and to compute e-ph scattering rates and the hole mobility. 
%
%
The band- and momentum-resolved e-ph scattering rates $\Gamma^{\textrm{e-ph}}_{n\textbf{k}}$ are obtained in the lowest order of perturbation theory \cite{Bernardi-review},
\begin{align}
\label{eq:e_scat_rate}
\Gamma^{\textrm{e-ph}}_{n\textbf{k}}=\;\frac{2\pi}{\hbar} & \sum_{m\nu\textbf{q}}|g_{nm\nu}(\textbf{k},\textbf{q})|^2\\[2pt]
\times\big[ & (N_{\nu\textbf{q}}+1-f_{m\textbf{k}+\textbf{q}})\delta(\varepsilon_{n\textbf{k}}-\varepsilon_{m\textbf{k}+\textbf{q}}-\hbar\omega_{\nu\textbf{q}})\nonumber\nonumber\\[5pt]
&+(N_{\nu\textbf{q}}+f_{m\textbf{k}+\textbf{q}})\delta(\varepsilon_{n\textbf{k}}-\varepsilon_{m\textbf{k}+\textbf{q}}+\hbar\omega_{\nu\textbf{q}})\big],\nonumber
\end{align}
where $\varepsilon_{n\textbf{k}}$ and $\hbar\omega_{\nu\textbf{q}}$ are the hole and phonon energies, respectively, and $f_{n\textbf{k}}$ and $N_{\nu\textbf{q}}$ the corresponding occupations. 
The scheme developed in our recent work \cite{Zhou2016} is applied to converge $\Gamma^{\textrm{e-ph}}_{n\textbf{k}}$. The relaxation times $\tau_{n\textbf{k}}$ used in the mobility calculations are the inverse of the scattering rates, $\tau_{n\textbf{k}}=1/\Gamma^{\textrm{e-ph}}_{n\textbf{k}}$. Our calculations focus on holes, and include only the HOMO and HOMO$-1$ bands because the energy gaps to the HOMO$-2$ and LUMO bands are larger than the highest phonon frequency.\\
%
%
\indent 
We employ the Boltzmann transport equation \cite{Zhou2016,Marzari2014_02} within the relaxation time approximation to calculate the electrical conductivity
\begin{eqnarray}
\label{eq:conductivity}
\sigma_{\alpha\beta} (T) =e^2\int_{-\infty}^{\infty}dE\left(-\frac{\partial f(E,T)}{\partial E}\right)\Sigma_{\alpha\beta}(E,T) 
\end{eqnarray}
where the transport distribution function $\Sigma_{\alpha\beta}(E,T)$ at energy $E$ and temperature $T$ is defined as
\begin{equation}
\label{eq:TDF}
\Sigma_{\alpha\beta}(E,T)=\frac{2}{V_{\textrm{uc}}}\sum_{n\textbf{k}}\tau_{n\textbf{k}}(T)v_{n\textbf{k},\alpha}v_{n\textbf{k},\beta} \, \delta (E-\varepsilon_{n\textbf{k}})
\end{equation}
and is calculated via tetrahedron integration \cite{Andersen1994}. The band velocities $\textbf{v}_{n\textbf{k}}$ are obtained from Wannier interpolation; $\alpha$ and $\beta$ are cartesian directions, and $V_{\textrm{uc}}$ is the unit cell volume. The hole mobility along the direction $\alpha$ is computed using $\mu_\alpha=\sigma_{\alpha \alpha}/n_{p}e$, where $n_{p}$ is the hole concentration. These e-ph and mobility calculations on unit cells with tens of atoms are made possible by efficient algorithms combining MPI plus OpenMP parallelizations we recently developed.\\%
%
%
\begin{figure*}[!th]
\includegraphics[scale=1.0]{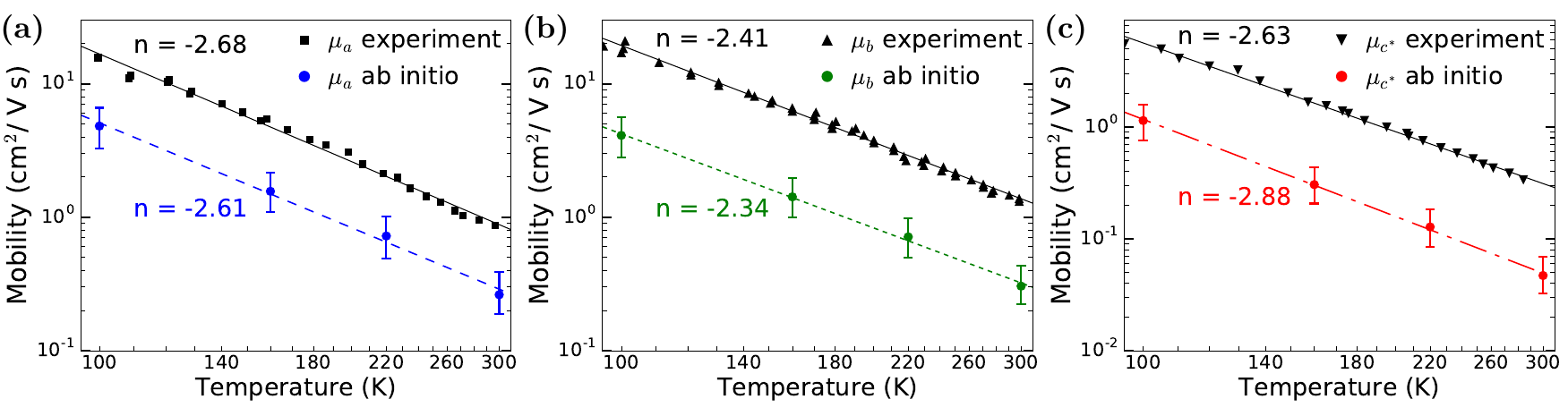}
\caption{The hole mobility in naphthalene, given, from left to right in separate panels, in the two in-plane directions \textit{a}, \textit{b} and in the plane-normal direction \textit{c}*. Circle markers are the computed mobilities and black markers the experimental data from Ref. [\onlinecite{Karl1985}]. Straight lines are best fits to the power law function $T^{-n}$ of the data points in the 100$-$300 K temperature range, and the exponent $n$ for each data set is also given. The error bars are obtained by assuming a $10\%$ error on both the phonon frequencies and the GW band stretching factor. These error sources are assumed to be independent and combined together.} 
\label{fig:mobility}
\end{figure*}
\indent
The computed bandstructures and phonon dispersions are given in the appendix (See Fig. \ref{fig:bands}). The GW correction is important as it stretches the valence band, thus lowering the hole effective mass and changing the relative alignment of the valence band valleys. The quality of our phonon dispersions is comparable with that of recent accurate phonon calculations in naphthalene \cite{Neaton2016}. For reference, we also employ the methods above to compute the phonon dispersion of the perdeuterated naphthalene. The comparison with experimental data is given in the appendix (See Fig. \ref{fig:D8}).\\
%
%
\section{results}
\begin{figure}[t]
\includegraphics{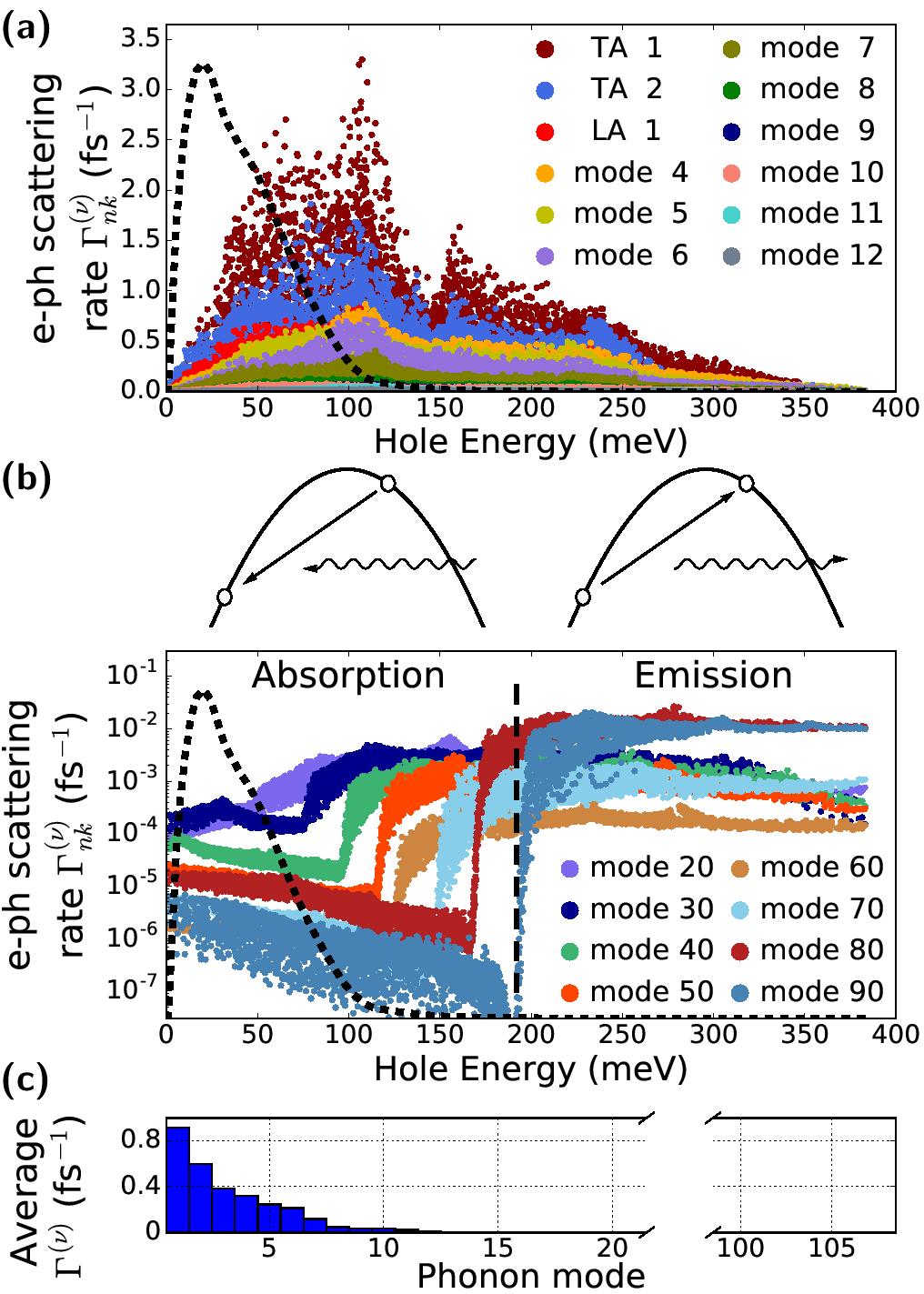}
\caption{Mode resolved e-ph scattering rates, $\Gamma^{(\nu)}_{n\textbf{k}}$, for (a) the 12 inter-molecular phonon modes and (b) selected intra-molecular phonon modes (note the y-axis log scale). Also sketched in (b) are the dominant e-ph scattering processes below and above the phonon emission threshold energy $\hbar \omega_0$, which is shown as a vertical dashed line for mode 90. The black dashed curve represents the integrand in Eq. (\ref{eq:conductivity}), and shows that only hole states within a 50$-$100 meV energy window of the valence band maximum (VBM) contribute to the mobility. (c) Mode-resolved scattering rates averaged over the energy window contributing to the mobility. In all plots, the zero of the energy axis is the VBM, and the hole energy increases moving away from the VBM into the valence band.} 
\label{fig:mode_analysis}
\vspace{-1.0\baselineskip}
\end{figure}

\vspace{-10pt}
\indent 
Figure \ref{fig:mobility} shows our calculated hole mobilities in the in-plane \textit{a} and \textit{b} and the plane-normal \textit{c}* directions (see Fig. \ref{fig:crystal_structure}). 
The experimental data given for comparison is taken from Ref. [\onlinecite{Karl1985}]. 
The computed mobilities are lower by a factor of 3 $-$ 5 than the experimental values; the smallest discrepancy (a factor of 3) is found for the \textit{a} direction, and the highest (a factor of 5) in the c* direction. 
Note that the \textit{c}* axis corresponds to a direction along which the molecules are stacked, so that the slightly lower accuracy in this direction is expected due to our neglect of van der Waals interactions in the e-ph coupling. 
%
%
Fitting the data with a power law function $T^{-n}$ over the 100$-$300 K temperature range yields calculated exponents $n$ in the 2.34$-$2.88 range for the three directions, in agreement \textit{within 3\%} (in the \textit{ab} plane) and 10\% in the \textit{c}* direction with the exponents $n$ obtained by fitting the experimental data (see Fig. \ref{fig:mobility}). 
The charge transport anisotropy is estimated by evaluating mobility ratios between different directions at 300 K. The computed ratios, $\mu_b/\mu_a=1.16$ and $\mu_{c\textrm{*}}/\mu_a=0.18$ are consistent with the experimental values of $1.57$ and $0.34$, respectively.\\
%
%
\indent
Since the accuracy of the phonon dispersions and GW bandstructures depends on the chosen crystal structure, exchange-correlation functional, and pseudopotential, 
it is important to quantify how these sources of uncertainty affect the computed mobility. To this end, we estimate how the combination of a small error in the GW correction (arbitrarily chosen to be $\sim$10\% in the stretching factor of the valence band) and an assumed $\sim$10\% error on the phonon frequencies (a conservative value for organic crystals) affect our calculations. The resulting error bars on the mobilities are given in Fig. \ref{fig:mobility}.\\ 
\indent
Within these uncertainties, which are typical of \textit{ab initio} methods $-$ especially for organic crystals with complex structures $-$ the range of computed mobilities (inclusive of the error bars) reaches values roughly 2$-$3 times smaller than the experimental result in the in-plane \textit{a} and \textit{b} directions. 
Overall, the temperature trends and absolute values of the mobility are remarkably accurate, particularly when compared to the very scarce literature on charge transport in organic crystals using \textit{ab initio} band theory. 
Our accuracy is comparable to the best calculations \cite{Negri2014,Troisi2007} using quantum chemistry methods based on hopping that dominate the literature.\\
\indent
We have verified that employing the Tkatchenko-Scheffler (TS) vdW correction \cite{TS1,TS2}, which is more accurate than the Grimme-vdW correction used here, does not change appreciably the structure and mobility. 
In particular, the root-mean-square (RMS) deviation between the atomic positions obtained with the Grimme-vdW and the TS-vdW corrections is only 0.05~{\AA}, and the RMS deviation of the bond lengths is $\sim$0.05\%. The mobility at 300 K obtained by computing the bandstructure, phonons and e-ph matrix elements with the structure obtained using the TS-vdW correction is very close (within 5$-$10\%, and thus within the error bars in Fig. \ref{fig:mobility}) to the mobility computed here using the Grimme-vdW method (see Fig. \ref{fig:TS_mobility} in Appendix). Future work will investigate further the role of the vdW correction on the e-ph coupling and mobility in organic crystals.\\
%
%
\indent 
Next, we investigate the role of different phonon modes in scattering the hole carriers. In the charge hopping picture, the conventional wisdom is that low-frequency inter-molecular phonon modes, which correspond to rigid motions of entire molecules \cite{Beljonne2013_02, Orlandi2006}, determine the mobility since they strongly affect the rate of charge hopping between molecules. Intra-molecular vibrations, on the other hand, are typically neglected due to their hypothesized weaker coupling to the carriers. 
There are 108 phonon modes in naphthalene, the 12 lowest-frequency modes are inter-molecular, and the others are intra-molecular. 
We express the total e-ph scattering rate in Eq. (\ref{eq:e_scat_rate}) as the sum of the scattering rates due to each individual mode $\nu$, i.e., $\Gamma^{\textrm{e-ph}}_{n\textbf{k}}=\sum_{\nu}\Gamma^{(\nu)}_{n\textbf{k}}$, and investigate the mode-resolved scattering rates $\Gamma^{(\nu)}_{n\textbf{k}}$. Here and in the following, the phonon modes are numbered in order of increasing energy at the Brillouin zone center, and the hole energy increases moving away from the valence band maximum (VBM) into the valence band.\\
\indent
Figure \ref{fig:mode_analysis}(a) shows the mode-resolved e-ph scattering rates as a function of hole energy for the 12 inter-molecular phonon modes, and Fig. \ref{fig:mode_analysis}(b) for selected intra-molecular phonons. 
Note that the inter-molecular phonons have either zero or very small minimum frequency since they correspond to transverse acoustic (TA) and longitudinal acoustic (LA) vibrations (modes 1$-$3) or other rigid vibrations or librations of the molecules (modes 4$-$12). By contrast, the intra-molecular modes 20$-$90 in Fig. \ref{fig:mode_analysis}(b) possess much higher frequencies. 
%
The integrand of the mobility in Eq. (\ref{eq:conductivity}) is also plotted in Figs. \ref{fig:mode_analysis}(a)$-$\ref{fig:mode_analysis}(b) to highlight the energy window contributing to the mobility, which spans hole states within 50$-$100 meV of the VBM. In this energy window, the 12 inter-molecular phonon modes exhibit much greater scattering rates than the intra-molecular modes, due to reasons related to the e-ph scattering phase space that are examined next.\\ 
\indent
%
%
In the hole scattering rates of Eq. (\ref{eq:e_scat_rate}), the first term in square brackets corresponds to phonon emission, and is proportional to the phonon population $N_{\nu\textbf{q}}$+1 since $f_{m\textbf{k}+\textbf{q}}\!\approx \!0$ for holes in our chosen temperature range. The term in the second square brackets is the phonon absorption rate, which is proportional to $N_{\nu\textbf{q}}$. 
Since the inter-molecular phonon modes 1$-$12 have a zero or small minimum energy, inter-molecular phonon absorption and emission processes are \textit{both} active at all hole energies. 
Their scattering rate decreases monotonically with phonon energy (and thus with mode number, since the modes are numbered in order of increasing energy). Similar to simple metals and non-polar inorganic semiconductors, the main source of scattering are acoustic modes, with smaller contributions from other molecular rigid vibrations and librations (modes 4$-$12). 
This result is further illustrated in Fig. \ref{fig:mode_analysis}(c), where the average $\Gamma^{(\nu)}_{n\textbf{k}}$ over the $100$ meV energy window of relevance for the mobility is given for each phonon mode. The dominant role of inter-molecular modes is consistent with the charge hopping intuition that rigid molecular vibrations mainly affect charge transport in organic materials. However, in our band picture based on phonon emission and absorption events, the origin of this behavior can be attributed to the phase space rather than the strength of the e-ph coupling per se, as further discussed below.\\
%
%
%
%
\indent 
The effect of intra-molecular phonons on the mobility is more subtle. Figure \ref{fig:mode_analysis}(b) shows that the e-ph scattering rates for these modes exhibit a trend with two plateaus as a function of hole energy. As explained next, the plateau at low hole energy corresponds to phonon absorption, and the one at higher hole energy to phonon emission. 
%
Consider an intra-molecular phonon with minimum energy $\hbar \omega_{0}$. Due to energy conservation, a hole in the valence band can emit such a phonon only at hole energy higher than $\hbar \omega_{0}$.  
At hole energies below this threshold, only phonon absorption is possible, with a rate proportional to the phonon occupation $ N_{\nu \textbf{q}} \!\propto\! e^{-\hbar \omega_0 / k_B T}$, which is much smaller than 1 at room temperature in naphthalene since most intra-molecular modes have minimum energies $\hbar \omega_0\!\approx\!$ 50$-$200 meV. 
Therefore, the plateau at hole energies below $\hbar \omega_0$ is associated with a small intra-molecular phonon absorption rate, and it spans the entire energy window contributing to the mobility.\\ 
\indent
At hole energies above $\hbar \omega_{\textrm{0}}$, the phase space for e-ph scattering increases dramatically since holes can emit intra-molecular phonons, with a rate proportional to $N_{\nu \textbf{q}} + 1$ and thus much greater than the absorption rate. 
Opening this phonon emission channel leads to an increase of the e-ph scattering rates by several orders of magnitude, but this increase occurs outside the energy window of relevance for charge transport due to the high energy of intra-molecular phonons in naphthalene. 
These trends are expected to be general in organic crystals, since the dominant presence of hydrogen, carbon and other light elements makes their intra-molecular phonon energies much greater than $k_B T$. Interestingly, in organic molecules containing heavy atoms, which introduce low-frequency intra-molecular vibrations, a contribution to transport from intra-molecular phonons is expected.\\
\indent
In short, the two-plateau structure for intra-molecular mode e-ph scattering is such that only the small rate for thermally activated phonon absorption falls in the energy range of interest for transport. Therefore the mobility is controlled by low-frequency inter-molecular vibrations. 
%
%
However, note that intra-molecular phonons are expected to dominate carrier dynamics at higher hole energy above the phonon emission threshold, where their combined scattering rate overwhelms that from the (much fewer) inter-molecular modes. 
This analysis shows that intra-molecular phonons play an essential role in the dynamics of excited carriers  \cite{Zhou2017,Zhou2016,Louie2014, Bernardi-review} in organic semiconductors.\\
%
%
\section{discussion}
\vspace{-10pt}
\indent 
While the phase space limits their scattering near the band edge, intra-molecular phonons can couple strongly with holes at all energies, and in fact more strongly than inter-molecular modes. To study this point, we compute the local e-ph coupling constants $g^{\textrm{(loc)}}_{\nu \textbf{q}}$ between each phonon mode at the Brillouin zone center ($\textbf{q}=0$) and the HOMO Wannier function (WF) $w_\textbf{R}$(\textbf{r}):
%
%
\begin{equation}
\label{eq:local_g}
g^{\textrm{(loc)}}_{\nu\textbf{q}}=\sqrt{\frac{\hbar}{2\omega_{\nu\textbf{q}}}}\langle w_\textbf{R} | \Delta_{\nu\textbf{q}}V^{\textrm{KS}}  | w_\textbf{R} \rangle,
\end{equation}
where $\textbf{R}$ is the WF center, and the change in Kohn-Sham potential $\Delta_{\nu\textbf{q}}V^{\textrm{KS}}$ arises from the atomic displacements $e_{\kappa\alpha,\nu}$ of each atom $\kappa$ (with mass $M_{\kappa}$) along all cartesian directions $\alpha$ due to the phonon mode $\nu$,\begin{equation}
\Delta_{\nu\textbf{q}}V^{\textrm{KS}} = e^{i\textbf{q}\cdot \textbf{r}} \sum_{\kappa\alpha} \frac{1}{\sqrt{M_\kappa}} e_{\kappa\alpha,\nu}\partial_{\kappa\alpha,\textbf{q}} V^{\textrm{KS}}.
\end{equation}
%
%
%
\begin{figure}[t]
\includegraphics{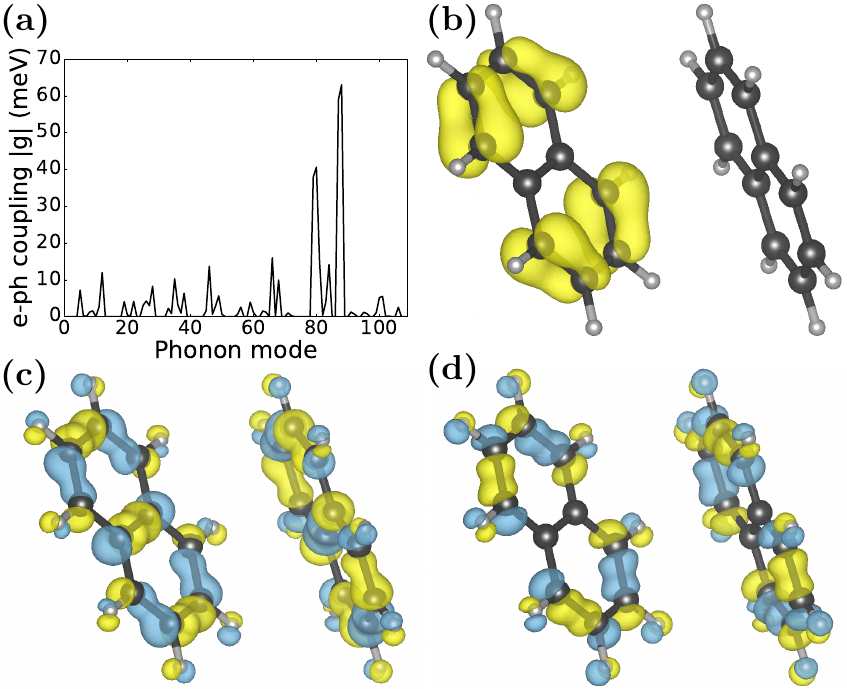}
\caption{(a) The absolute value of the local coupling constant [see Eq. (\ref{eq:local_g})] between each of the phonon modes and the HOMO Wannier function. (b) The square of the HOMO Wannier function. The potential perturbation $\Delta_{\nu\textbf{q}}V^{\textrm{KS}}$ at $\textbf{q}=0$ is shown for (c) mode $\nu=88$ and (d) mode $\nu=89$. These modes correspond to the peak (mode 88) and sudden drop (mode 89) in e-ph coupling in (a). In panels (b)$-$(d), yellow is used for positive, and blue for negative isosurfaces.}
\label{fig:g_analysis}
\vspace{-1\baselineskip}
\end{figure}
%
%
\indent
The absolute value of these local e-ph coupling constants are shown in Fig. \ref{fig:g_analysis}(a) for all 108 phonon modes \footnote{Note that the e-ph coupling constants for the three acoustic phonon modes vanish as $\textbf{q}\rightarrow 0$}. 
Contrary to intuition, the strongest e-ph coupling to the HOMO hole state is not with the inter-molecular modes that control transport. 
Rather, specific high-frequency intra-molecular phonons (in particular, modes 79$-$88) exhibit the strongest coupling to holes. To understand this result, we plot quantities entering the local e-ph coupling in Eq. (\ref{eq:local_g}), namely the square of the HOMO WF, $|w_\textbf{R}(\textbf{r})|^2$, and the perturbation potential $\Delta_{\nu\textbf{q}}V^{\textrm{KS}}$ due to the atomic motions associated with the given mode.\\
\indent 
Figure \ref{fig:g_analysis}(b) shows the square of the HOMO WF orbital, $|w_\textbf{R}(\textbf{r})|^2$; the perturbation potential $\Delta_{\nu\textbf{q}}V^{\textrm{KS}}(\mathbf{r})$ at $\textbf{q}=0$ is shown in Fig. \ref{fig:g_analysis}(c) for mode 88 and Fig. \ref{fig:g_analysis}(d) for mode 89, which are respectively cases of maximally strong and weak e-ph coupling. We find that e-ph coupling is maximal for mode 88 due to the strong overlap between the square of the HOMO WF and the perturbation potential, and the fact that both quantities possess the same sign over most of the molecule, so that no cancellations occur in the real-space integral in Eq. (\ref{eq:local_g}). 
By contrast, the symmetry of mode 89 is such that its perturbation potential $\Delta_{\nu\textbf{q}}V^{\textrm{KS}}(\textbf{r})$ alternates positive and negative lobes at bonds where the square of the HOMO WF is large. As a result, the \textit{integrand} $\left | w_\textbf{R} (\textbf{r}) \right|^2 \cdot \Delta_{\nu\textbf{q}}V^{\textrm{KS}}(\textbf{r})$ in Eq.~\ref{eq:local_g} is positive for two bonds and negative (and roughly equal in absolute value) for the other two bonds, thus leading to a small integral over the entire molecule in Eq.~\ref{eq:local_g}. This cancellation results in a small e-ph coupling for mode 89. Other phonon modes are either associated with perturbation potentials with small overlap with the square of the HOMO WF, as is the case for modes in which only the hydrogen atoms vibrate, or with perturbations that are out of phase with the square of the HOMO WF, similar to the case of mode 89. 
This analysis shows that the atomic displacements and mode symmetry critically determine the e-ph coupling of intra-molecular modes, which can be much stronger than that of inter-molecular modes due to the large spatial overlap between the square of the HOMO WF and the intra-molecular mode perturbation.\\
%
%
\indent 
Lastly, we comment on the fact that our computed phonon-limited mobility is smaller than the experimental result. Due to the presence of impurities and defects in real samples, our calculation is expected to provide an upper bound to the mobility, and thus to slightly overestimate its experimental value, consistent with our recent results for inorganic crystals \cite{Zhou2016}. 
The reason why our result is lower than experiment is unclear, but a possible cause is the neglect of non-adiabatic effects.\\
\indent
Our method employs only the lowest Born--Oppenheimer potential energy surface (PES), since the e-ph perturbation potential is computed using DFPT. However, an insight from non-adiabatic surface hopping calculations \cite{BlumbergerReview,Tavernelli2013} is that several PESs can lie close in energy in organic crystals, and including their contributions to charge transport may increase the mobility. 
The impact of such non-adiabatic effects on the mobility within the band theory framework used here deserves further investigation. Nonetheless, the fact that our results underestimate the measured mobility is important as it further supports the conclusion in Ref. [\onlinecite{Stojanovic2012}] that hole charge carriers in naphthalene crystals are weakly coupled to phonons, so that transport occurs in the band-like regime studied here. In fact, polaronic effects resulting from strong e-ph coupling (beyond the lowest order employed here) would further suppress carrier transport by increasing the scattering rates and effective masses \cite{Mahan2000}, thus reducing the mobility.\\
%
%
\section{conclusion}
\vspace{-10pt}
\indent 
In summary, we compute with quantitative accuracy the hole mobility and its temperature dependence in naphthalene, dramatically improving the agreement with experiment compared to previous efforts using band theory to study charge transport in organic crystals. 
%
Our results show that \textit{ab initio} approaches based on band theory and many-body perturbation theory are well equipped to compute charge transport in organic semiconductors. They can provide an accuracy at least as satisfactory as widespread quantum chemistry methods based on charge hopping, as well as insight into the role of different phonon modes. 
%
Our work sets the stage for attempting higher-order corrections or diagram resummations in the e-ph perturbation to access the strong e-ph coupling regime typical of polaron transport.
%
\section{ACKNOWLEDGMENTS}
\begin{acknowledgments}
The authors thank Maurizia Palummo for discussions. 
N.-E.L. acknowledges the Physics department at Caltech for the TA Relief Fellowship. M.B. and L.A. acknowledge support by the National Science Foundation under Grant ACI-1642443, which provided for basic theory and electron-phonon code development. J.-J. Zhou was supported by the Joint Center for Artificial Photosynthesis, a DOE Energy Innovation Hub, as follows: the development of the scattering rate and mobility calculations was supported through the Office of Science of the U.S. Department of Energy under Award No. DE-SC0004993. This research used resources of the National Energy Research Scientific Computing Center, a DOE Office of Science User Facility supported by the Office of Science of the U.S. Department of Energy under Contract No. DE-AC02-05CH11231.
\end{acknowledgments}
%
\vspace{-1\baselineskip}
\appendix*
\section*{APPENDIX}
\vspace{-2\baselineskip}
\begin{figure}[!th]
\includegraphics{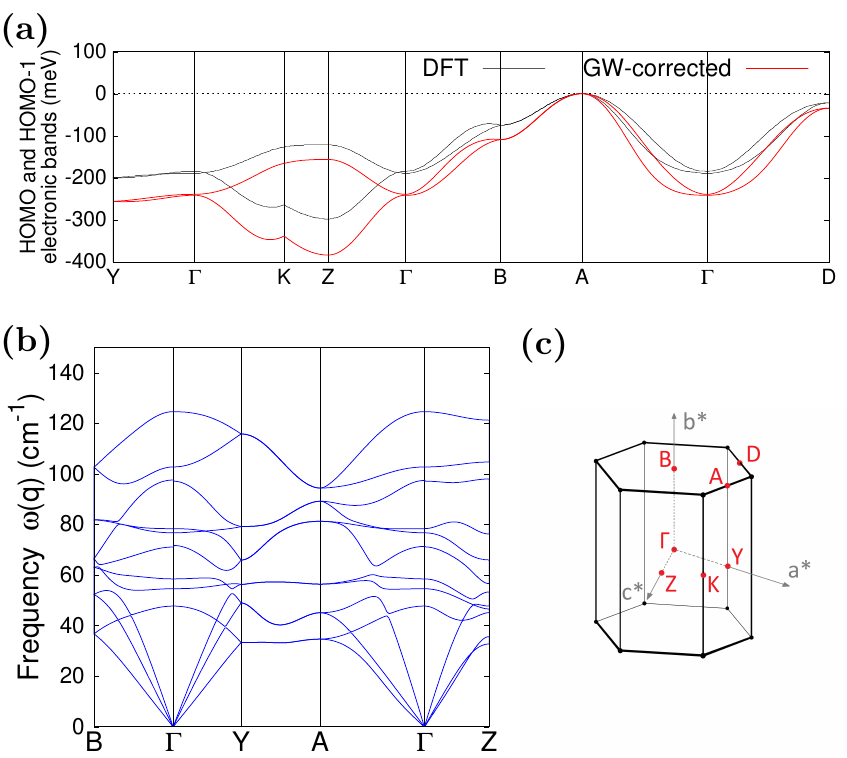}
\caption{Bandstructures and phonon dispersions of naphthalene crystal, for the structure used at 300 K. (a) The HOMO and HOMO$-1$ electronic bands, where black is used for the DFT bands and red for the bands with the GW correction. (b) Dispersion of the 12 inter-molecular phonon modes. (c) Sketch of the first Brillouin zone.}
\label{fig:bands}
\end{figure}
\begin{figure*}[!th]
\includegraphics[scale=1.0]{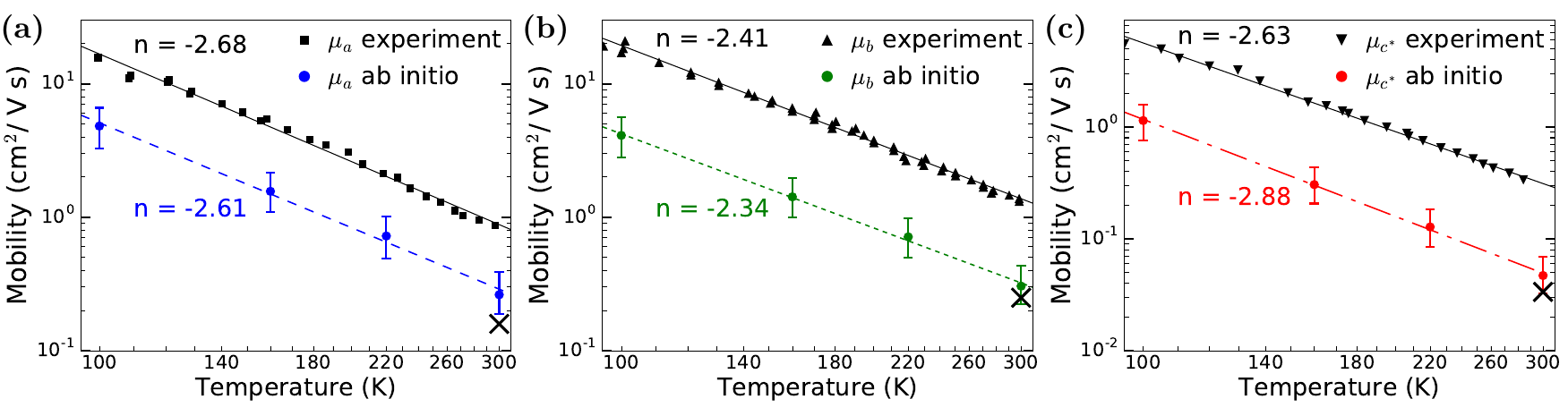}
\caption{The mobility at 300 K obtained using a structure relaxed with the TS-vdW correction is shown with black crosses. The values fall within the error bars.}
\label{fig:TS_mobility}
\end{figure*}
\begin{figure}[!th]
\includegraphics{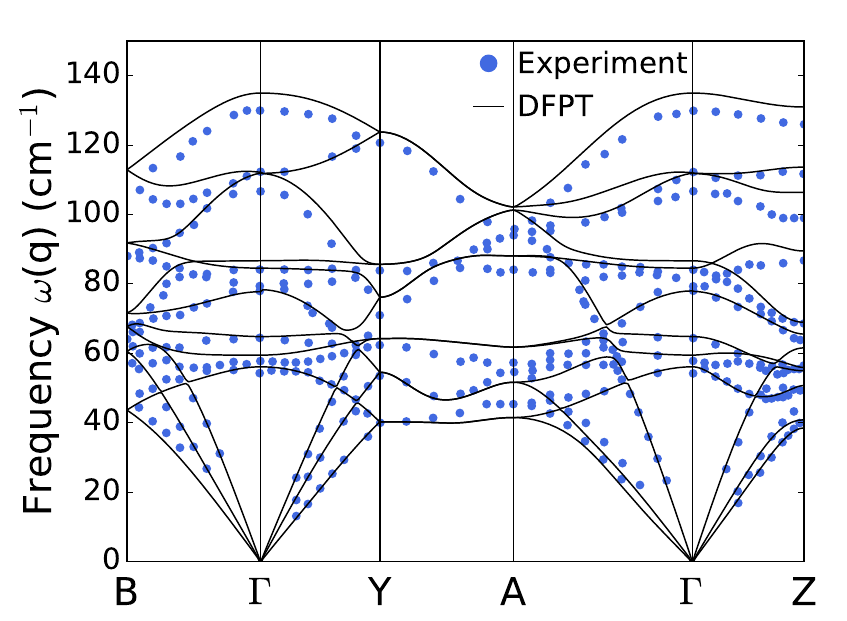}
\caption{Calculated dispersion of the 12 inter-molecular phonon modes for perdeuterated naphthalene, with lattice constants taken from Refs. [\onlinecite{webCSD},\onlinecite{Neaton2016}]. The markers are the experimental data at 6 K from Ref. [\onlinecite{D8phonon}].}
\label{fig:D8}
\end{figure}
\bibliographystyle{apsrev4-1}
\bibliography{naphthalene_mobility}

\end{document}